# SpecSwap-RMC: A novel reverse Monte Carlo approach using a discrete configuration space and pre-computed properties

*Mikael Leetmaa, Kjartan Thor Wikfeldt, and Lars G.M. Pettersson[*]*

FYSIKUM, Stockholm University, AlbaNova University Center, SE-106 91 Stockholm, Sweden

E-mail: lgm@fysik.su.se

**Abstract**

We present a novel approach to reverse Monte Carlo (RMC) modeling, SpecSwap-RMC, which makes use of pre-computed property data from a discrete configuration space replacing atomistic moves with swap moves of contributions to a sample-set representing the average, or summed property. The approach is particularly suitable for disordered systems and properties which require significant computer time to compute. We demonstrate the approach by fitting jointly and separately the EXAFS signal and x-ray absorption spectrum (XAS) of ice Ih using as SpecSwap sample-set 80 configurations from a space of 1382 local structures with associated pre-computed spectra. As an additional demonstration we compare SpecSwap and FEFFIT fits of EXAFS data on crystalline copper finding excellent agreement.

**Introduction**

The Reverse Monte-Carlo (RMC) method has been available as a tool for modeling diffraction data for over 20 years now [1]. The possibility to build three-dimensional structure models consistent with the measured data has had an enormous impact on how we understand the structure of disordered materials. RMC is indeed a powerful tool for structure modeling, but although the algorithm in its original formulation is quite general and could be used to model any kind of data that derive from a geometrical structure, it has seldom been used for other purposes than to model diffraction[2], EXAFS [3] and various geometrical constraints[4-6]. The reasons for this limited range of data types have been accuracy and speed. For use in the conventional RMC algorithm a structural probe must be evaluated fast enough to be performed for each trial move, and at the same time be described accurately enough to give reliable structural information. Only few structural probes fulfill these criteria.

To overcome these obstacles on the road to realizing the full potential of RMC, we here devise a new RMC version, SpecSwap-RMC, which makes use of pre-computed property data from a discrete configuration space. A large set of structures is chosen (the basis set), for instance from an RMC structure model or an MD simulation. All relevant properties from the structures are computed and collected in a library. A subset of the basis (the sample set) is taken out to represent the state of the system. A swap move is then defined that propagates the state of the system with the fit to a reference property as driving force in an RMC-like fashion. Full details of the procedure will be given below. With steadily increasing computer power this approach of pre-computing properties for a large set of atomic configurations and using the collected data in an RMC scheme becomes quite appealing. Once all desired properties from a set of structures are computed, which can be done, *e.g.*, on a massively parallel system without severe restrictions on computational time, the RMC procedure itself becomes extremely fast.

In the following the underlying theory will be detailed and the applicability demonstrated by, as a test case, modeling the experimental near-edge x-ray absorption spectrum (XAS) and the EXAFS signal of ice Ih using a library of 1382 pre-computed XAS spectra and EXAFS signals using structures from a path-integral Molecular Dynamics (PI-MD) simulation. The discussion and characterization of the detailed structures generated by the fit and their significance for the understanding of the details in the spectroscopies applied to ice will be given in a separate publication[7].

Another test case, the simulation of fcc Cu metal EXAFS, will be provided as a comparison between EXAFS analysis performed with SpecSwap-RMC on the one hand and the well established FEFFIT fitting method [8] on the other. In our method, crystalline Cu clusters with imposed random disorder along with their calculated EXAFS oscillations provide the basis set. Since Cu metal can be accurately characterized with the combination of multiple-scattering FEFF [9] calculations and subsequent FEFFIT fits, the satisfactory outcome of the methodological comparison with results from SpecSwap-RMC is highly encouraging. It points to the feasibility of SpecSwap as a method for structure modeling of EXAFS data, particularly for disordered materials and in cases when other datasets or geometrical constraints can be included.

**Theory**

We will begin by outlining a completely general theory of RMC which will clarify the characteristic features of the approach. To this end we define a set *B* with *N* elements

$$B \equiv B_N \equiv \{x_N\} \qquad (1)$$

We shall call the set *B* the basis set. Define also a set *S*

$$S_M \subset B_N \qquad (2)$$

selected from *B*, with $M << N$. This set will be referred to as the sample set.

Introduce now a move operator $\hat{m}$, defined by its action on *S*

$$\hat{m}S_M \equiv S'_M \qquad (3)$$

where $S' \subset B$, selected from *B* by some rule depending on *S*.

Let *S* represent the state of some system. Any property *P(S)* of the system depending on (the elements of) *S* will potentially differ before and after the application of $\hat{m}$. Define an error function $\chi^2$, as the squared difference between the property *P(S)* and a reference property *R*, divided by a parameter $\sigma^2$ related to the assumed uncertainty of the reference.

$$\chi^2 = \frac{(P(S) - R)^2}{\sigma^2} \qquad (4)$$

The extension to more than one property is trivial:

$$\chi^2 = \sum_i \chi_i^2 = \sum_i \frac{(P_i(S) - R_i)^2}{\sigma_i^2} \qquad (5)$$

The change in $\chi^2$ due to the application of $\hat{m}$ is then the sum over contributions $\Delta\chi_i^2$ from each property

$$\Delta\chi_i^2 = \frac{1}{\sigma_i^2}\left[(P_i(\hat{m}S) - R_i)^2 - (P_i(S) - R_i)^2\right] \qquad (6)$$

We now have what is needed to perform a (reverse) Monte-Carlo (MC) simulation on the system, by successive applications of $\hat{m}$ to *S*, always accepting the change from $S_M$ to $S'_M = \hat{m}S_M$ if $\Delta\chi^2 \leq 0$ and otherwise accepting the change if

$$rnd[0,1] < \exp(-\Delta\chi^2) \qquad (7)$$

where *rnd*[0,1] denotes a random number between 0 and 1. If the change is rejected, the new state of the system *S'* is taken equal to the previous state *S* (in accordance with the Metropolis Monte-Carlo (MMC) algorithm [10]).

So far we have not done much more than restated the original RMC method [1] in slightly more general terms. However, written in this way the concept behind RMC becomes very clear, while all to us known versions of RMC conform to this generalization. The original RMC method [1] is obtained if we let *B* contain the set of all possible coordinate configurations of some box of atomic coordinates (which makes the number of elements $N$ in $B_N$ practically infinite), let $S$ contain one such coordinate configuration (thus $M=1$) representing the state of the system, let the rule for making a move $\hat{m}$ be defined as "move some randomly selected particle a short random distance", and thus replacing $S$ with another set $S' \subset B$ according to a rule depending on (the elements of) $S$, and let the properties $\{P(S)\}$ be, *e.g.*, the diffraction pattern, or some other property which can be computed from the coordinates of the atoms (*i.e.* from the state $S$ of the system). We note that in the GNXAS RMC version of Di Cicco and Trapananti [11] a large set of molecular replica (*i.e.* a large number of elements in $S$) was used to successfully describe the distribution of distances of a $Br_2$ molecule in a fit to the measured EXAFS signal. The basis set $B$ can in that case formally be regarded as the set of all possible Br-Br distances, as they are given by the resolution of the particular machine representation.

For an MC simulation to perform an efficient sampling of the configuration space $\Delta\chi^2$ should, when equilibrium is reached in the simulation, be small enough to allow a substantial fraction of the attempted moves [12]. This means that $\hat{m}$ must be defined so that the new state of the system $S'$ is close enough to the previous state $S$; thus the overlap between successive states should be large. If $M >> 1$ this can be obtained by defining $\hat{m}$ to let $S$ and $S'$ differ by one element only. If $M$ is small or equal to one (as in the original RMC) this is obtained by letting $\hat{m}$ replace the single element $x_i \in S$ with another element $x_j \in B$, which has large overlap with $x_i$. In practice this is, in most MC simulations of liquids, done by letting $\hat{m}$ choose a new set of coordinates $x_j$ differing from $x_i$ in the position of one or a few atoms only and then limit the size of the step until the desired ratio of accepted over rejected moves is obtained. Another possibility, mostly suited for amorphous covalent solids, is to let the move swap the positions of two atoms of different elements, as implemented in the latest version of the RMC++ code [4].

Different acceptance criteria (equation (7)) have been proposed for use in RMC based on the sampling of the configuration space compared to MMC simulations of a Lennard-Jones liquid [13,14]. It is of course straightforward in the formalism outlined above to replace the acceptance criterion eq. (7) with some other criterion of choice. The standard MMC-like test should however be enough to consider for the present discussion.

*The SpecSwap-RMC Method*

We will now sketch another version of the generalized RMC described above, which for the basis set $B$ uses a finite set of $N$ small clusters of atomic coordinates (with radius ~10 Å) each representing an atom or molecule with its immediate surroundings in a liquid or solid. It should be noted that this limitation on the type of structure used is only for convenience in the present implementation and is not a theoretical limitation of the method. One could just as well use periodic boxes of coordinates or replicas of a gas phase molecule in different geometric conformations as elements in $B$.

We take a large sample set $S_M$ for which still $1 << M << N$. $\hat{m}$ is then defined as: "pick a new set $S' \subset B$ which differs from $S$ in only one element". In this way no move in the conventional sense (of moving the coordinates of an atom) is made, but instead an element $x_i \in S$ is replaced with another element $x_j \in B$ chosen at random. We will then rely on pre-computed, tabulated, properties to determine $\Delta\chi^2$ for the swap-move according to equation (7).

The most fundamental difference between SpecSwap-RMC and the original RMC, except from the purely technical (yet quite important) issues of using pre-computed properties in SpecSwap, is the size of the basis *B* itself. In any RMC version that makes a move by changing atomic positions the size of *B* is practically infinite (limited by the machine resolution and the size of the atomic system), while in SpecSwap a finite set of a few thousand to perhaps tens of thousand structures can be used. We are in this case limited by the cost to pre-compute and store the properties for each element $x \in B$. The SpecSwap approach thus allows for extremely fast evaluation of the RMC loop, at the cost of a much smaller configuration space (as defined by *B* and the size of *S*).

It is thus likely that the results of the fit will be strongly dependent on the choice of basis set *B*. Therefore, to judge the quality and meaning of a fit, it is necessary to relate the obtained solution to the basis in a well-defined way. (This is similar to the EPSR method [15-17] where results ultimately should be related to the choice of initial potential.) For this purpose we devise the following weighting method: Once $\chi^2$ is converged we start recording how often each element in *B* appears in *S*. Normalized to the number of times we probed *S* we obtain weights $w_i$ which converge for each $x_i \in B$ during the simulation. It is then possible to take the weighted sum of each property *P* from each element $x \in B$

$$P_B^w = \frac{1}{N} \sum_i^N w_i P(x_i) \qquad (8)$$

and the property without weights

$$P_B = \frac{1}{N} \sum_i^N P(x_i) \qquad (9)$$

The meaningful comparison, as the result of the SpecSwap-RMC simulation, is then to look at how $P_B^w$ differs from $P_B$.

It could also be illustrative to sort the basis elements according to decreasing weights and sum together properties from different chunks of the sorted list. In this way it is possible to investigate, *e.g.*, what type of structural motifs are most important in the basis set for the description of a certain reference, and how their properties differ from elements with low weight. Note that a large weight not only indicates that the basis element was indeed important for the description of the property, but also that that type of basis element was underrepresented in the basis set, *i.e.* it could not easily be replaced by another basis element. A low weight, on the other hand, does not necessarily signify that the basis element was unimportant. It could well be, but it could also be due to that type of basis element being represented sufficiently well in the basis set such that it could easily be replaced by another member of the same category. The weights thus give the proper weighting of the contributions from the library for the fit to the data and, as such, indicate the relative importance of different basis elements, *i.e.* how the basis or underlying model can be improved.

As a final theoretical consideration before we go on to the case studies demonstrating the concepts let us look at how the SpecSwap method can be used to investigate information overlap between different properties. Each element $x \in B$ consists of a set of coordinates $c$, together with $N_P$ properties $p_1, p_2, ..., p_{N_P}$ derived from the coordinates.

$$x_k \equiv \left\{ c^k, \left\{ p_1^k, p_2^k, ..., p_{N_P}^k \right\} \right\} \qquad (10)$$

We can fit one property, say $P_a = \frac{1}{M} \sum_i^M p_a^i$, to its reference $R_a$ while comparing another property $P_b$ with its reference $R_b$ without fitting the latter. The coincidence of $P_b$ with $R_b$ will then depend on the extent of information overlap between the properties $P_a$ and $P_b$. Remember, however, that in SpecSwap this should always be related to the given basis set $B$ as described with the weighting procedure above. The concept of information overlap of course applies to any RMC method. If one property is fitted to an experimental reference whilst another is not, the latter will generally come out well compared against its reference only if there is essential information overlap between the properties.

**Application to XAS and EXAFS of ice Ih**

As an example application of the SpecSwap-RMC method we fit the near-edge X-ray Absorption Spectrum (XAS) of bulk ice Ih from ref. [18] and the EXAFS signal of bulk ice from ref. [19,20]; the fits will be done separately as well as in combination.

The basis set was compiled from 1382 local structures from a single snapshot of a path-integral molecular dynamics (PIMD) simulation performed at 200K with 1523 water molecules using the SPC/E potential [21]. The XAS signal was computed from each basis element structure represented as a cluster with 39 water molecules centered around the selected molecule; the transition-potential approach[22] as implemented in the StoBe-deMon DFT code[23] was used to generate the spectra. Each computed spectrum was computationally energy-calibrated as described in ref.[24] and a flat 0.5 eV Gaussian broadening was used on the discrete oscillator strengths. Full details regarding the XAS calculations can be found in ref.[24]. For each of the 1382 calculated spectra a total of approximately 12 hours CPU time was required for the three steps (ground state, transition potential and variationally determined core-excited state[25]) to obtain a spectrum; this is clearly beyond reach of any implementation of RMC using atomistic moves.

As mentioned in the introduction, EXAFS has earlier been implemented in RMC methods based on atomistic moves, *e.g.* in the RMCA[1,3], and the latest RMC++ [4] codes, as described in ref.[3]. However, due to the requirement of rapid evaluation of the signal for each atomistic move these implementations only include the single-scattering contribution through the pair-correlation function $g(r)$ according to equation (11) below. The GNXAS approach of Di Cicco and co-workers [26] has been extended to combine fits to EXAFS and $g(r)$ functions from diffraction in an RMC fashion. The RMC-GNXAS approach [11] is, however, also limited to fitting explicitly only the two-body scattering contribution, even though the contribution from three-body correlations can be extensively analyzed within the same framework after the fit [27]. A proper description of the EXAFS signal from water and ice for use in an RMC simulation, however, requires a multiple-scattering approach to account for the focusing effect of hydrogen due to which a phase-shift occurs and the scattering from oxygens with strong hydrogen-bonds becomes significantly enhanced[19,20,28]. We have therefore used the FEFF8.4 code[9] to compute the full signal of each of the 1382 local PIMD ice structures. It was found that an SCF radius of 4.0 Å and a path-expansion radius of 8.0 Å gave converged results; note that this implies larger clusters than used for the XAS calculations. Due to an inherent problem in FEFF with short interatomic bonds, the intramolecular O—H contribution to EXAFS was omitted; this introduces only minor errors since the intramolecular contribution is very small. Finally, the adjustable $\Delta E_0$ parameter in FEFF was optimized beforehand to give a good overall EXAFS phase of the basis set. This parameter will be allowed to vary during the SpecSwap-RMC simulation in future versions for general applicability to EXAFS structure modeling.

In Figure 1a we first compare the sum of all XAS spectra in the basis set with the experimental spectrum; it is clear that there are significant differences between the summed spectrum and experiment. In the following we shall assume that the calculated spectrum for each structure is sufficiently accurate to provide a reliable prediction of its spectral contribution. The difference between the summed spectrum and the experiment can then only be due to the underlying structure from the simulation not being a true representation of the experimental structure. The PIMD structures used in the basis set can be regarded as variations around a perfect ice crystal, however the force-field itself provides only an approximate description of the interactions. The experiment could on the other hand have contributions from defects, grain boundaries, surface effects etc. The *ansatz* is then that enough different local structures are present in the basis set for the SpecSwap-RMC method to come up with a re-weighted structural solution in better agreement with the local structure of the experimental ice.

In Figure 1b we show representative individual spectra in the basis and compare with the overall sum. It is clear that not one individual spectrum will resemble the overall sum and the final shape of the computed spectrum must then be the result of an averaging process; the goal of SpecSwap-RMC is then to provide proper weights in the summation for the given basis.

In Figures 2a and 2b we show the corresponding comparisons for the computed EXAFS data using the same basis set. From Figure 2a we see that the unweighted summation of all contributions from the basis *B* already gives a rather good representation of the EXAFS reference-data from ref. [19,20]. Again, as seen in Figure 2b, the individual contributions show very large variations in both amplitude and phase and should thus constitute a potentially good basis for a SpecSwap-RMC simulation.

Having thus characterized the basis set based on the PIMD ice simulation, and stressed the need for the present SpecSwap-RMC approach with regards to both time required to compute the property (XAS) and the need to include relevant physical effects (EXAFS), we now use SpecSwap-RMC on this basis to investigate whether a different weighting of the elements can be found to reproduce the experimental data. We furthermore perform the simulations for each data set separately as well as simultaneously, targeting both experimental data sets. In all cases the basis *B* includes all 1382 structures and the sample set *S* contains 80 elements. Weights are recorded each 1000 accepted moves making sure that the fit was fully converged well before beginning to record weights.

The SpecSwap-RMC simulation of the XAS data was performed during 900,000,000 attempted moves, with a total of 1,241,959 accepted moves and consequently weights recorded 1241 times; it should be emphasized again that a move here is defined as replacement of one randomly selected spectrum in the sample set with a randomly selected spectrum from the basis set without any shifts or modifications of the spectra. To conform to the XAS sum rule the spectrum from the sample set was area normalized to the experimental spectrum up to 545 eV before each calculation of $\Delta\chi^2$, where $\Delta\chi^2$ was computed according to equation (6).

Figure 3a shows the XAS spectrum of the basis set (green), and the XAS spectrum weighted according to the converged weights from the simulation (red). The reference spectrum is also shown (black). It is immediately obvious that the SpecSwap-RMC method has been successful at fitting the reference curve using the given basis. The weighted curve is in very close resemblance with the reference, while the curve from the basis set without weights is completely lacking the characteristic sharp pre- and main-edge features at 535 and 537 eV indicating that structures giving features in these energy regions are present but underrepresented in the basis set.

We turn now to the EXAFS signal from the structures. Figure 3b shows the summed EXAFS signal from the basis set (green), and the signal multiplied with the weights from the fit to XAS (red). Although the weighting changes the XAS signal quite a lot (figure 3a), the EXAFS signal is almost completely unchanged. This is a consequence of the different structural sensitivities between the near-edge and extended part of the x-ray absorption spectrum. EXAFS is sensitive mostly to pair correlations and H-bond angles while the near-edge XAS spectrum is sensitive to various higher-order correlations [29]; there is thus not much structural information overlap between the two data sets.

We now run the SpecSwap-RMC simulation against the EXAFS data using the EXAFS signals computed from the same 1382 PIMD structures. To balance the emphasis of the fit over the entire k-range the fit is performed to the fine-structure $\chi(k)$ multiplied by $k^2$. The simulation was run during 100,000,000 attempted moves with 1,228,989 accepted moves and the weights were thus collected 1228 times in this case. The resulting fit is shown in figure 4a with the summed EXAFS signal from the basis set (green), and the signal multiplied with the weights from the fit to the data (red) together with the experimental target data (black). Even though the agreement is rather good for the unweighted basis set, we see that the phase is improved at higher k values in the weighted sum, indicating that the SpecSwap run has indeed been selective in terms of changing the balance between contributions. On the other hand, applying the weights determined from the EXAFS data to a summation of the computed XAS spectra in the basis set (figure 4b) shows very little difference from the unweighted sum, except for a slight shift to higher energy of the broad post-edge peak around 540-541 eV in the XAS spectrum. This additionally underlines the complementarity of the two experimental probes; the significance of the observed shift in terms of structure and the detailed properties of the spectroscopic versus scattering experimental probes will be discussed in a separate publication[7].

Having ensured that it is indeed possible to fit the XAS and the EXAFS reference data separately and having established that the two data sets do contain complementary structural information, we now attempt to fit them both simultaneously using the same basis set, *i.e.* the 1382 PIMD structures. The simulation was run during 900,000,000 attempted moves with 268,593 accepted moves and weights were thus collected 268 times. One should ultimately make a longer run to obtain better statistics in the weights, however the weights converged quite fast, which makes the present sampling sufficient to ensure reliable resulting trends. However, it is clear that the requirement to simultaneously fit both data sets makes the acceptance rate significantly lower which again underlines the lack of information overlap between the data sets.

The resulting simultaneous fits to both XAS and EXAFS data are shown in figures 5a and b from which it is clear that the procedure has been successful in finding a set of weights for this basis such that both the XAS and EXAFS data may be rather accurately reproduced. We emphasize the fact that the SpecSwap-RMC procedure only explores a limited, predetermined phase space given by the basis set $B$ and the size of $S$, but within this phase space it has the capability to produce a set of weights for a much improved representation of the data.

Each individual XAS and EXAFS contribution is associated with a particular local structure. The resulting weights from the SpecSwap-RMC simulation applied to the structures in the basis set can be viewed as an adjustment of the relative occurrence of the different local structures in the basis set to produce an improved fit to the experimental data. As in any other RMC approach, uniqueness of the solution in terms of structure is by no means guaranteed and will depend strongly on the structure-sensitivity of the property being fitted and, in the case of SpecSwap-RMC, also on the basis set provided.

Figure 6 shows the weights per basis element in the combined fit sorted according to increasing weight and normalized to the occurrence in the sample set. A basis element with the weight 50% thus means it has been found in the sample set 50% of the times when weights have been recorded. We see from the figure that some basis elements are much more likely to be found in the sample set than others. This illustrates that SpecSwap-RMC as a technique is able to identify important contributions that are underrepresented in the basis set and thus have a much higher probability of being present in the sample set when the statistics are taken; this can then form the basis for a detailed analysis of the experimental data in terms of underlying atomistic structure. This will be investigated in some depth for the case of XAS and EXAFS on ice Ih in a separate publication[7].

**Application to Cu metal EXAFS**

Cu metal can be accurately characterized with FEFFIT fits to experimental EXAFS data [30]. Hence, Cu provides an important test case for SpecSwap-RMC modeling of EXAFS. Here, EXAFS data on Cu at 50 K, as provided with the IFEFFIT package[31], have been used.

We will for simplicity limit our methodological comparison to the first-shell scattering contribution. In FEFFIT this implies fitting the first peak of $\tilde{\chi}(r)$, *i.e.* the Fourier transform of the experimental $\chi(k)$, and from SpecSwap-RMC we need the first shell contribution to $g(r)$ which is subsequently used together with back-scattering amplitudes calculated with FEFF8.4 to construct the first-shell contribution in $k$-space:

$$\chi(k) = 4\pi\rho \int r^2 g(r) \chi_{pair}(k,r) dr, \qquad (11)$$

where $\rho$ is the atomic number density, $\chi_{pair}(k,r)$ is the back-scattering amplitude and $g(r)$ is in our case restricted to only the first peak. In this way, the total first-shell scattering contribution of both methods can be directly compared in $k$-space.

The conventional FEFFIT fits are straight-forward. A scattering path from the first shell is retrieved from a FEFF8.4 calculation on crystalline Cu and used in FEFFIT for a determination of the $E_0$ shift and the first-shell distance, along with its amplitude reduction and Debye-Waller factors. These fits are performed in $r$-space where the first shell contribution can easily be separated; the resulting fit is nearly perfect.

The SpecSwap-RMC simulations were performed with a basis set composed of 1000 clusters of Cu atoms, each cluster containing random displacements from the known crystalline lattice and an overall random distance scaling factor. 40 basis elements were used in the sample set, and out of 100,000,000 attempted moves around 2 million were accepted.

Prior to the simulation, the EXAFS signal was calculated for each cluster with FEFF8.4 using an SCF radius of 4.5 Å and maximum path-length of 11.2 Å. These parameters provide almost converged results for EXAFS on Cu which is entirely sufficient for a methodological comparison of the first-shell contribution. An important consideration at this stage concerns the choice of $E_0$ shift in the EXAFS calculations. Here, the $\Delta E_0$ derived from FEFFIT was used, but to rid the SpecSwap-RMC method of its dependence on model systems or FEFFIT fits a built-in optimization of the $\Delta E_0$ parameter should be used. This can be incorporated in SpecSwap-RMC and will be added in future versions.

There is a fundamental difference between how the two methods are applied here; in FEFFIT the isolated first-shell contribution is fitted in $r$-space, while in SpecSwap-RMC the entire experimental dataset is fitted in $k$-space. For specific applications it may be advantageous to construct the basis set instead from the Fourier transformed data, $\tilde{\chi}(r)$, from each basis element, and fit to the experimental $\tilde{\chi}(r)$ shell-by-shell. However, with applications to disordered materials in mind, where the shell-by-shell procedure is inappropriate, we will here illustrate the SpecSwap-RMC method by fitting the total EXAFS data in $k$-space.

Clusters with considerably larger disorder, *i.e.* with larger random distortions of the crystalline structure, were also investigated. Fitting a basis set composed only of these clusters in SpecSwap-RMC turned out to give poor results; the amplitude was far from that of the experimental Cu EXAFS data, illustrating that most basis functions had canceling contributions to the EXAFS oscillations. This demonstrates the importance of a good quality basis set where realistic structures are not completely absent.

We show in Figure 7a the result from the SpecSwap-RMC fit against EXAFS on Cu. The curve from the basis set is shown together with the resulting weighted curve. Figure 7b shows resulting $g(r)$, weighted and unweighted. A good agreement with the data is reached, and the *g(r)* is rather strongly affected by the fit. The small remaining misfits of the experimental EXAFS curve can be attributed to the finite range of structures in the basis set as well as to the truncation of higher-shell scatterings in the individual basis element EXAFS calculations. For a comparison of the first-shell scattering however, going to even larger clusters or extending the basis set should not be necessary. This is particularly true since the first shell in Cu is well separated from higher shells, which also leads to a clear separation of frequency components in *k*-space.

Figure 8 compares the first-shell contribution to EXAFS from SpecSwap-RMC and FEFFIT. The unfitted contributions from each method differ significantly from each other, but the converged fits are in very good agreement. Identical results should not be expected, primarily due to the inherent finiteness of the configurational phase-space in SpecSwap-RMC.. Also, SpecSwap-RMC does not optimize the agreement with data in a least-squares sense, but rather samples a much broader region of phase space where the agreement is close to but not exactly at the global minimum in $\chi^2$ (*i.e.* in the deviation from the data). In analogy with the conventional RMC method, this should be seen as an advantage of SpecSwap-RMC, in particular because of possible uncertainties in all experimental datasets and theoretical calculations. Comparing finally the derived distances of the two methods, we find that FEFFIT yields a first shell distance of 2.547 Å while SpecSwap, when a gaussian is fitted to the first Cu-Cu RDF peak, gives 2.549 Å. The difference of 0.002 Å is within the estimated statistical error of the FEFFIT fit.

**Conclusions**

We have outlined a new RMC version, SpecSwap-RMC, which operates on a large, but discrete, configuration space. No moves of atomic coordinates are made. Instead, a library of pre-computed properties for a large set of fixed geometries is used, and a move is defined as a change in the subset of geometries taken to represent the physical system. Evaluating $\Delta\chi^2$ for the move (equation (6)) then simply amounts to summing properties for different subsets of the full library. In this way most of the CPU-time consumption is moved outside the RMC-loop, making the SpecSwap-RMC simulation itself extremely fast.

The general applicability of SpecSwap-RMC to EXAFS structure analysis was demonstrated by the successful fits for both ice and copper and the good agreement with results from the FEFFIT method. In particular for highly disordered systems will the method provide significant advantages in EXAFS studies. Perhaps its main strength, however, is the straightforward inclusion of practically any property that depends on structure and can be calculated theoretically (*without* any severe restrictions on computational speed). This provides an enormous generalization of EXAFS and multiple dataset analysis, and is applicable to ordered as well as disordered materials.

We have in the case of fitting XAS of ice Ih, for the first time seen a full TP-DFT XAS calculation being used in an RMC-like scheme. This opens up many new possibilities. There are numerous experimental techniques available, *e.g.*, NMR, IR/Raman, x-ray Compton scattering, and core-level spectroscopies such as XAS, XES (x-ray emission spectroscopy), EXAFS, and XPS (x-ray photoemission spectroscopy), that provide structural information about a system, but require time-consuming quantum chemical calculations for their theoretical description[8,9,22,32-38]. It is clear that the method described in this paper will be a useful tool for data analysis and structure modeling in cases such as these, where a proper theoretical evaluation of the structural probe is far too time-consuming to be used in a conventional RMC simulation. Fully aware that the configuration space is not extensively sampled we still expect that our method will be a help in extracting useful structural information; especially from data that would otherwise be modeled with one or a few local geometries only, as has often been the case when studying, *e.g.*, (near-edge) XAS.

We note finally, from equation (10), that for the algorithm to work there is no need for a reference to an underlying structure. When modeling a physical atomistic system the reference to an underlying structure is *crucial* for the correct interpretation of the results. However, what the present algorithm needs is only a set of associated properties for each element $x \in B$. This characteristic might well open up for the use of the SpecSwap-RMC algorithm in fields quite different from the modeling of physical atomistic systems.

**Acknowledgements**

We thank Alexander Lyubartsev for providing the PI-MD ice structure. This work was supported by the Swedish Research Council (VR). Generous grants of CPU time at the Swedish NSC and HPC2N computer centers made this study possible.

**Figure Captions**.

**Figure 1**: **a)** XAS spectrum from the basis set taken as the average of all 1382 individual spectra (green) and experimental reference curve from ref. [18] (black). **b)** Average XAS spectrum from the basis set (black) and seven typical individual basis element spectra. (colour online)

**Figure 2**: **a**) EXAFS signal from the basis set taken as the average of all 1382 individual signals (green) and experimental reference curve from ref. [19] (black). **b)** Average EXAFS signal from the basis set (black) and seven typical individual signals. (colour online)

**Figure 3**: Results from SpecSwap-RMC run against XAS reference data [18]. **a)** Reference XAS spectrum (black), spectrum from library (green) and resulting weighted spectrum (red). **b)** EXAFS signal from library (green) and weighted signal using weights from the fit to XAS (*i.e*. not fitted to EXAFS) (red). The experimental EXAFS signal from ref. [19] (black) is shown for comparison. (colour online)

**Figure 4**: Results from SpecSwap-RMC run against EXAFS reference data [19]. **a)** Reference EXAFS signal (black), signal from library (green) and resulting weighted signal (red). **b)** XAS spectrum from library (green) and weighted spectrum using weights from the fit to EXAFS (*i.e*. not fitted to XAS) (red). The experimental XAS spectrum from ref. [18] (black) is shown for comparison. (colour online)

**Figure 5**: Results from SpecSwap-RMC run against both XAS and EXAFS reference data. **a)** Reference XAS spectrum (black), spectrum from library (green) and resulting weighted spectrum (red). **b)** Reference EXAFS signal (black), EXAFS signal from library (green) and weighted signal (red). (colour online)

**Figure 6**: Weights per basis function (percent occurrence in the sample-set). (colour online)

**Figure 7**: **a)** SpecSwap-RMC fit to Cu EXAFS reference from [31]. Resulting weighted curve (red) and curve from the basis set (green). **b)** Resulting change in g(*r*), weighted curve (red) and unweighted (green). (colour online)

**Figure 8**: Comparison of first-shell contribution from SpecSwap-RMC and FEFFIT fits. Resulting weighted SpecSwap curve (red) and curve from the basis set (green); FEFFIT starting point (grey) and resulting FEFFIT curve (blue). (colour online)

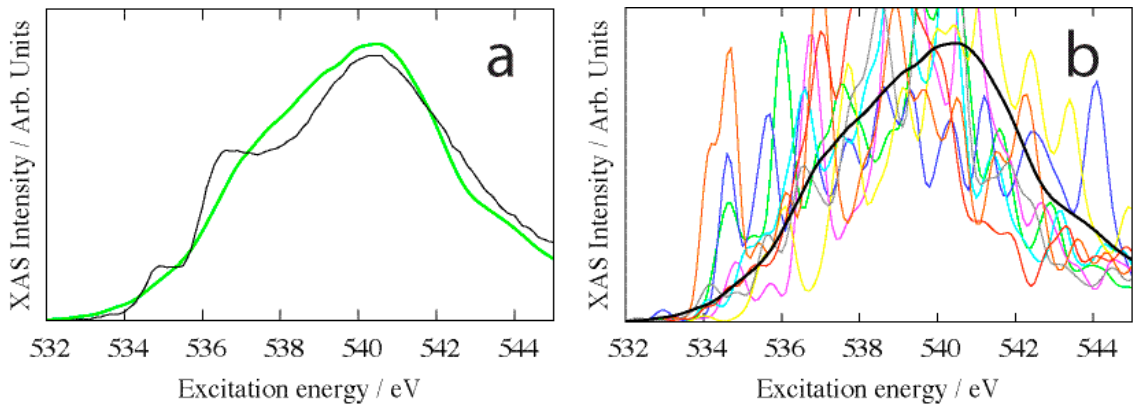

**Leetmaa *et al*, Figure 1**

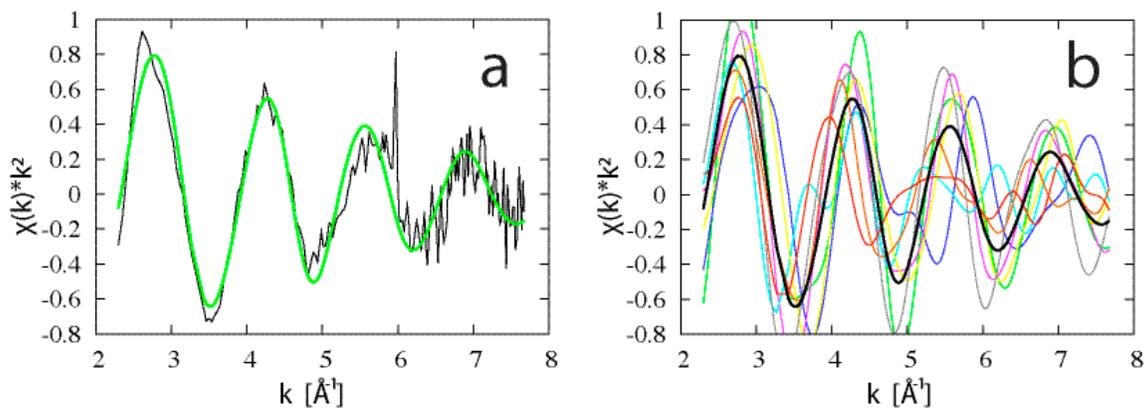

**Leetmaa *et al*, Figure 2**

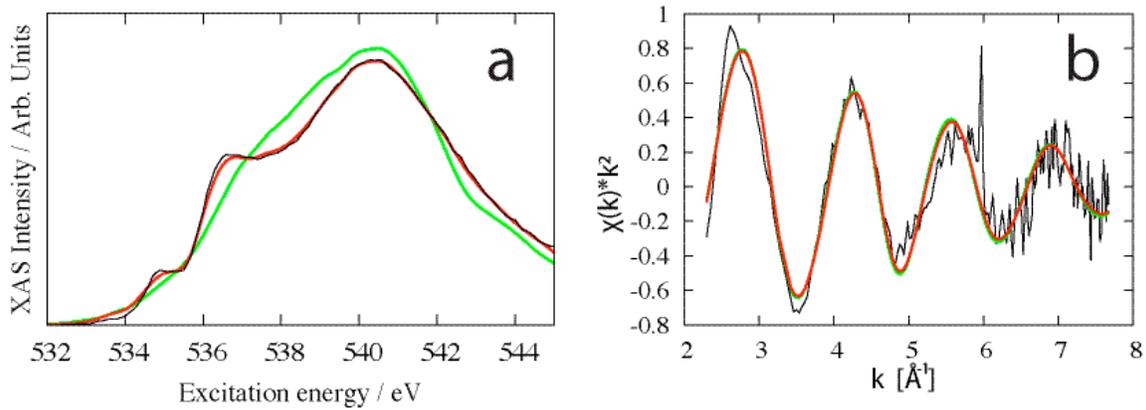

**Leetmaa *et al*, Figure 3**

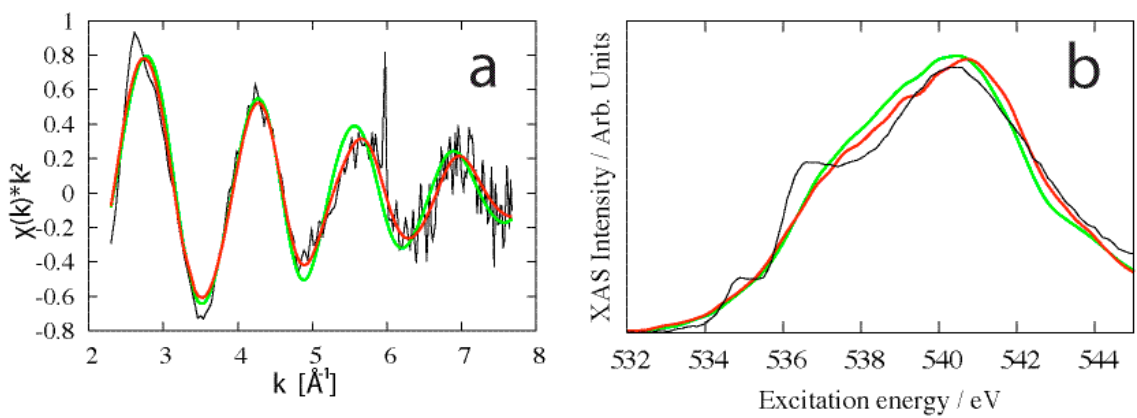

**Leetmaa *et al*, Figure 4**

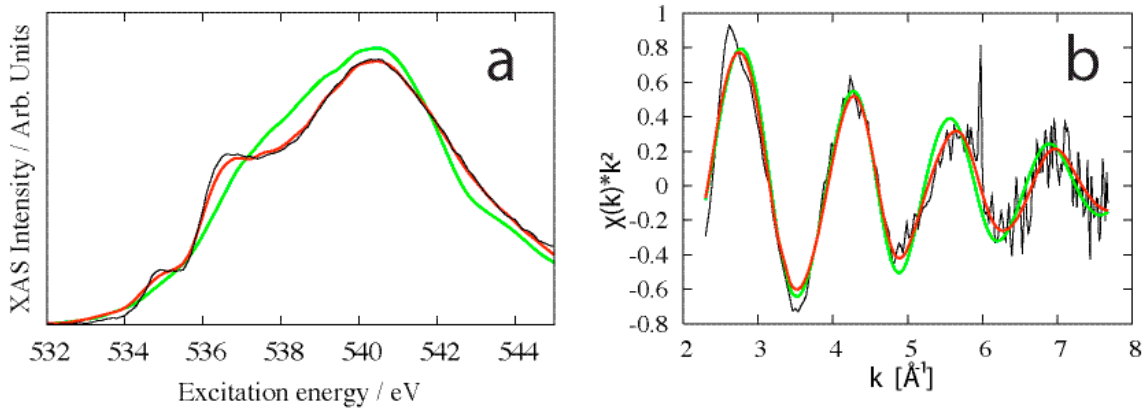

**Leetmaa *et al*, Figure 5**

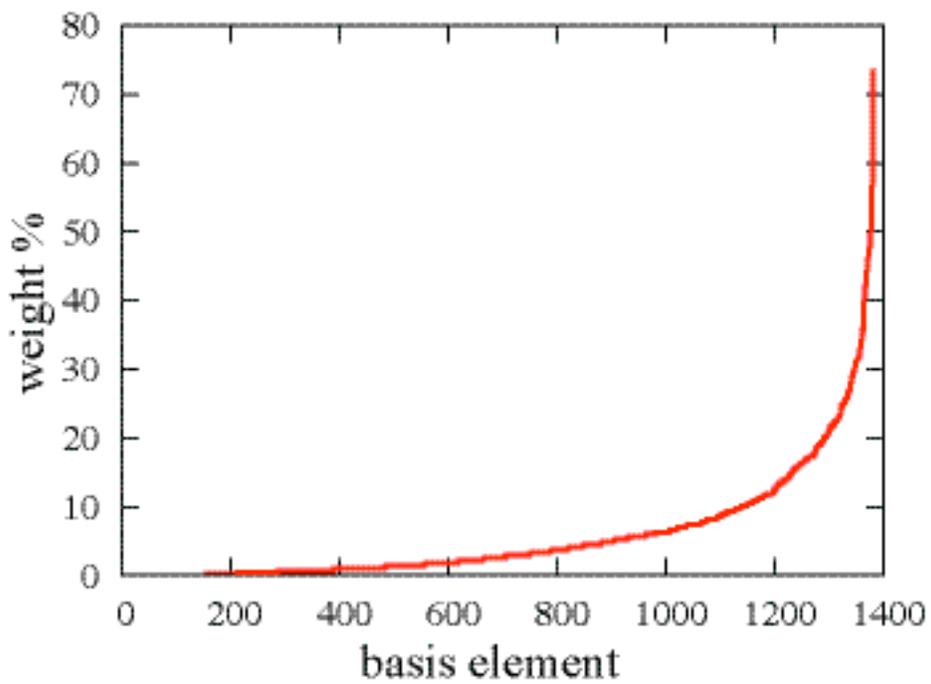

**Leetmaa *et al*, Figure 6**

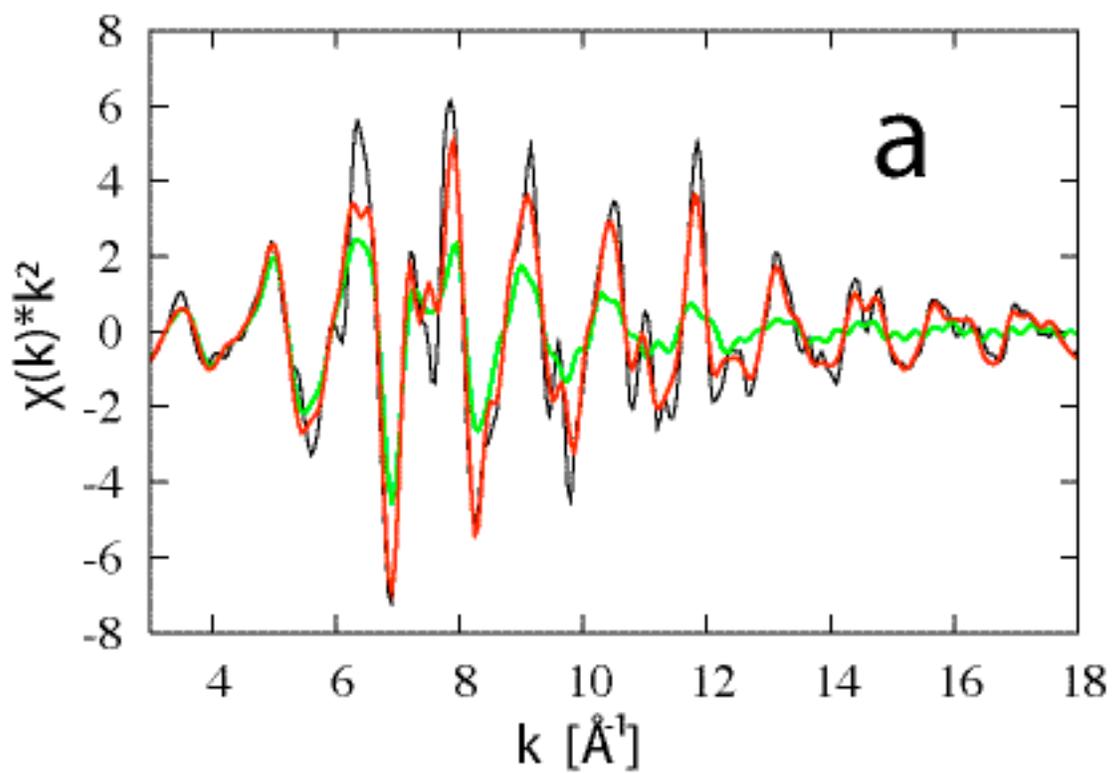

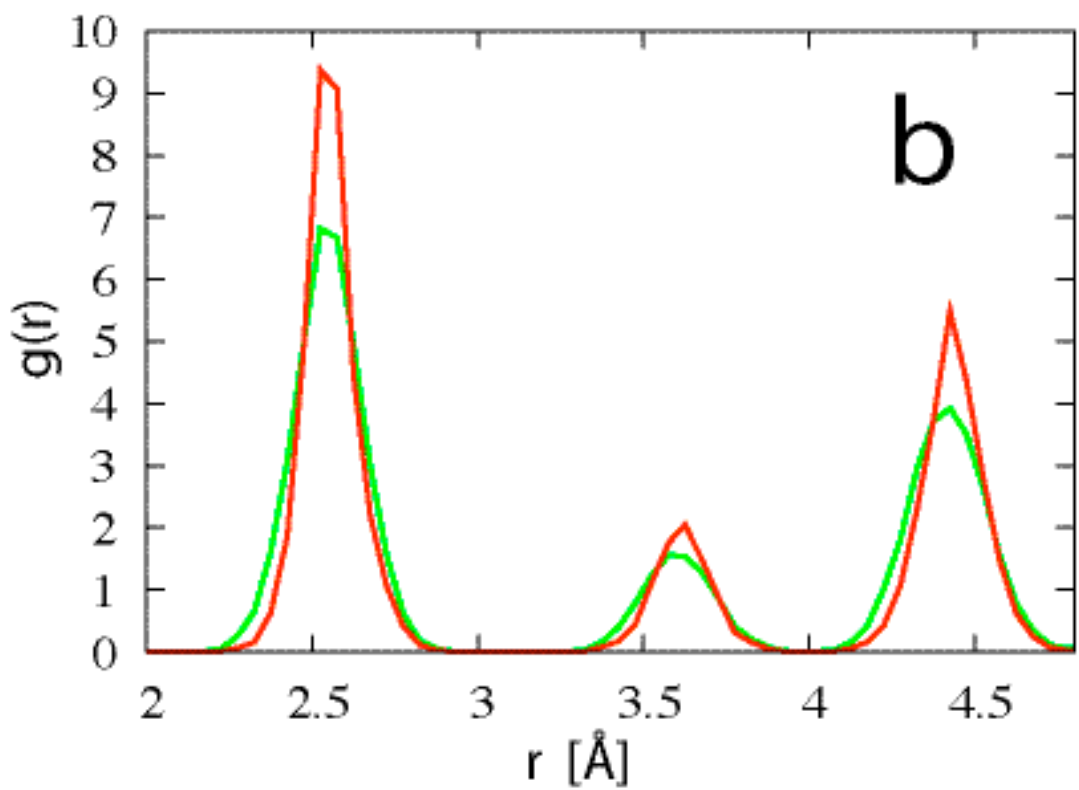

**Leetmaa *et al*, Figure 7**

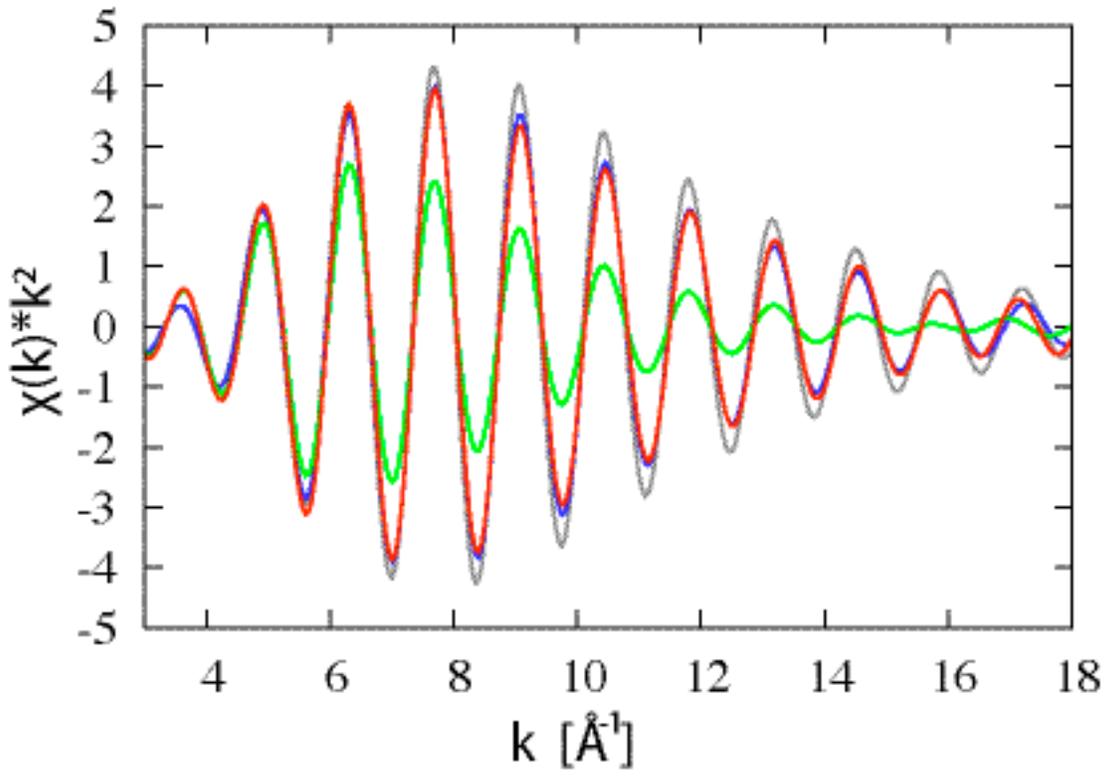

**Leetmaa *et al*, Figure 8**